%% file: main.tex
\documentclass[conference]{IEEEtran}
\IEEEoverridecommandlockouts

\usepackage{cite}
\usepackage{amsmath,amssymb,amsfonts}
\usepackage{graphicx}
\usepackage{textcomp}
\usepackage{xcolor}
\def\BibTeX{{\rm B\kern-.05em{\sc i\kern-.025em b}\kern-.08em
    T\kern-.1667em\lower.7ex\hbox{E}\kern-.125emX}}

\usepackage[T1]{fontenc}
\usepackage{url}
\usepackage{xspace}
\usepackage[acronym,nomain,nonumberlist]{glossaries}

\newacronym{ai}{AI}{artificial intelligence}
\newacronym{ci}{CI}{continuous integration}
\newacronym{bsp}{BSP}{bulk synchronous parallel}
\newacronym{mpi}{MPI}{Message Passing Interface}
\newacronym{hpc}{HPC}{high performance computing}
\newacronym{aws}{AWS}{Amazon Web Services}
\newacronym{gke}{GKE}{Google Kubernetes Engine}
\newacronym{aks}{AKS}{Azure Kubernetes Service}
\newacronym{eks}{EKS}{Elastic Kubernetes Service}
\newacronym{aiml}{AI/ML}{Artificial Intelligence and Machine Learning}
\newacronym{ml}{ML}{machine learning}
\newacronym{rdma}{RDMA}{Remote Direct Memory Access}
\newacronym{os}{OS}{operating systems}
\newacronym{vm}{VM}{virtual machine}
\newacronym{llm}{LLM}{Large Language Models}
\newacronym{fom}{FOM}{Figure of Merit}
\newacronym{sdk}{SDK}{Software Development Kit}
\newacronym{efa}{EFA}{Elastic Fabric Adapter}
\newacronym{ec2}{EC2}{Elastic Compute Cloud}
\newacronym{ucx}{UCX}{Unified Communication X}
\newacronym{cni}{CNI}{container networking interface}
\newacronym{ebpf}{eBPF}{extended Berkeley Packet Filter}
\newacronym{gvnic}{gVNIC}{Google Virtual NIC}
\newacronym{uri}{URI}{unique resource identifier}
\newacronym{api}{API}{application programming interface}
\newacronym{dag}{DAG}{directed acylcic graph}
\newacronym{crd}{CRD}{custom resource definition}
\newacronym{mcp}{MCP}{Model Context Protocol}
\newacronym{hci}{HCI}{Human-Computer Interaction}
\newacronym{gpu}{GPU}{Graphical Processing Unit}
\newacronym{qpu}{QPU}{Quantum Processing Unit}
\newacronym{qrmi}{QRMI}{Quantum Resource Management Interface}
\newacronym{pid}{PID}{Process IDentifier}

\begin{document}

\title{Hybrid Quantum and Classical Workload Management with Graph-based Scheduling}


\author{\IEEEauthorblockN{Vanessa Sochat}
\IEEEauthorblockA{\textit{Lawrence Livermore National Laboratory} \\
Livermore, California, USA \\
sochat1@llnl.gov \\
ORCID 0000-0002-4387-3819}
\and
\IEEEauthorblockN{Daniel Milroy}
\IEEEauthorblockA{\textit{Lawrence Livermore National Laboratory} \\
 Livermore, California, USA \\
 milroy1@llnl.gov \\
 ORCID 0000-0001-6500-3227}
 }

\maketitle

\input{sections/abstract}


\input{sections/1-introduction}
\input{sections/2-methods}
\input{sections/3-results}
\input{sections/4-discussion}

\section*{Acknowledgment}
 We are grateful for beaches and mountains, purple flowers, and our ability to
learn and grow. We thank the QRMI community for their discussions and
 collaboration. This work was performed under the auspices of the U.S.
 Department of Energy by Lawrence Livermore National Laboratory under Contract DE-AC52-07NA27344 (LLNL-CONF-2021259).

\bibliographystyle{IEEEtran}
\bibliography{references}

\end{document}

%% file: sections/abstract.tex
\begin{abstract}
High Performance Computing (HPC) centers are expanding to integrate quantum resources, enabling hybrid quantum-classical workflows for complex optimization. Integrating quantum processing units (QPUs) into workload managers poses an orchestration challenge: a remote QPU introduces a second queue---a ``two-queue problem''---alongside the scheduler's own. We present Fluence, a Kubernetes scheduler plugin backed by the Fluxion graph-based scheduler, enabling gang-scheduled placement for quantum-classical workloads and custom resources. First, under contention, Fluence's atomic gang placement eliminates the node-time a default scheduler wastes on partially placed gangs. Second, a synchronization primitive gates consumers behind a single producer's shared quantum task, cutting worker idle time roughly 1.2-12x under short queues and orders of magnitude under long ones. Third, policy-aware backend selection cuts mean per-run cost roughly 72x and time-to-result from hours to under two minutes. Together, these results show that quantum-awareness can be added to a cloud-native scheduler without modifying user containers.
\end{abstract}

%% file: sections/1-introduction.tex
\section{Introduction}

The \gls{hpc} center of the future is an autonomous and hybrid environment for scientific discovery, provisioning resources for scientific workloads that cross environments. A heavily accelerating technology is quantum computing, with applications moving from experimental to real world use-cases \cite{quantum-tech-review}. In the same way that a user can request accelerators such as \gls{gpu} devices, quantum devices will be added to the resource graph in the \gls{hpc} center of the future. 
To mitigate issues with noise \cite{Preskill2018-vw}, research has adopted a hybrid quantum-classical framework that encompasses a group of algorithms called Variational Quantum Algorithms (VQAs) \cite{Cerezo2021-cq}. Under this paradigm, classical resources are used to handle error minimization and parameter updates, and quantum devices are handed discrete tasks to work on. In simple terms, a classical process constructs a parameterized quantum circuit, submits it to a \gls{qpu} for execution, and receives a set of measurement outcomes for further interpretation. The circuit's parameters can then be adjusted and the process repeated until a defined convergence is reached. Canonical examples of this approach include the Quantum Approximate Optimization Algorithm (QAOA) \cite{Farhi2014-mw} and the Variational Quantum Eigensolver (VQE) \cite{Peruzzo2014-bh} that target combinatorial optimization and ground-state energy estimation, respectively.

From a workload scheduling standpoint, a quantum job is a classical job with a quantum dependency. VQA workloads primarily need quantum and classical resources to be available together to permit problem construction through post-processing via communications between the two systems. Since quantum resources can be accessed via a remote \gls{api} or allocated via a second queue, synchronization becomes a challenge. We call this the two-queue problem, which generalizes to scheduling to two different systems and manifests as a resource management challenge that traditional managers were not initially designed to solve. 

\smallskip
\noindent{\bf Models of Quantum Resource Management} 
Models of quantum resource management are characterized by different degrees of transparency and control. An institution directly managing a quantum device retains control and ability to see the state of the device. An institution that interacts with a remote quantum device has limited visibility about state and often interacts with the device via a middleware interface. A co-scheduling strategy of quantum devices with classical resources will depend on the visibility and ability to manage the devices. Having visibility at the finest granularity of scheduling detail allows greater ability to control temporal allocation and schedule for multi-tenancy use cases. A work request to a workload manager that directly manages quantum and classical resources can be received as a single job in one queue. 

A second queue is introduced in the case of using a remote \gls{api} service such as
Amazon Braket \cite{braket} or IBM Quantum Cloud \cite{IBM_Quantum_2026}. The same workload manager is required to interact with a second queue and co-schedule classical and quantum resources. The visibility of device state is unknown, and the only information provided by a vendor \gls{api} is the queue depth. Under this resource management model, the scheduler must attempt to co-schedule based only on the queue depth. An abstraction such as a reservation that provides a guarantee about executing \emph{now} or an \emph{<exact time in the future>} can provide enough transparency for a workload manager to account for in its scheduling model. This two-queue problem is further exacerbated by intermittent availability of the machines themselves, as the devices often need to be calibrated. This fluctuation that leads to downtime has been coined the ``offline problem'' \cite{Fang2025-wn}.

\smallskip
\noindent{\bf Workload Managers Integrating Quantum} 
A workload manager that attempts quantum integration must deal with the pathology of two uncoordinated queues. One of the most comprehensive efforts is a collaboration between IBM, Pasqal, STFC Hartree Centre, and Rensselaer Polytechnic Institute built around two layered components: the \gls{qrmi} and Slurm SPANK plugins that consume it \cite{Bacher2025-hi}. \gls{qrmi} is designed to be a vendor-agnostic middleware layer that makes it easy to access, control, and monitor quantum systems. 

An accompanying SPANK plugin allows for integration into the Slurm workload manager, which models a quantum device as a \gls{qpu} Slurm Generic Resource (GRES).  The plugin controls acquisition, execution, and release of quantum devices via the Slurm prolog, task initialization, and the epilog, respectively. The user selects a specific backend name, but the scheduler cannot currently choose a backend for a job based on dynamic criteria or properties. The plugin  ensures that the remote vendor \gls{api} is working, allows the job to continue, and then serves as a wrapper to export environment variables from a system configuration file. Since the session is requested in the Slurm prolog, the classical resources are already dispatched and must wait for the work to move up to the top of the queue.



The study that introduces the SPANK plugin~\cite{Bacher2025-hi} does not consider the cost of idle classical resources. We identify and quantify this two-queue problem --- classical worker pods idling on allocated nodes while the quantum task is queued externally --- and show that gating the classical workers until the quantum task is about to execute reclaims the worker idle time. Accomplishing this task requires a scheduling-layer that is orthogonal to the resource-control interface that \gls{qrmi} provides. 

A second concern is the need to store credentials within the cluster administrator configuration. The benefit of this approach is to facilitate group access to a shared credential. The design makes the assumption that all users share the same remote resources, and accounting must be handled by Slurm. While precautions can be taken to ensure protection of these secrets, their existence at the level of the system scheduler can introduce more security concerns than user-space credentials. For example, if the credentials are used in the job environment during \emph{slurmstepd}, this means that they would be available under a process environment and easily accessed by a user. Such access would allow users to not only view or control other users' jobs in the queue, but potentially access results. 


Quantum resources can also be represented in Kubernetes. The Qubernetes (Q8s) project extends Kubernetes by providing custom resource definitions for quantum resources, and defines quantum-capable nodes and jobs to schedule with a quantum aware scheduler. Pods are thus scheduled onto quantum-capable nodes that are identified via labels and that advertise their \gls{qpu} capacity \cite{Stirbu2024-lp}.  The project targets isolated quantum tasks and does not address multi-stage hybrid workflows, co-scheduling, backend selection, or contention with other jobs on the cluster. From the view of the scheduler, quantum resources are not modeled as distinct from classical nodes with labels.

A second project in Kubernetes takes an integrated approach of using existing components to orchestrate quantum work. Tejedor et al. present a cloud-native framework for quantum-HPC pipelines that use Argo Workflows for workflow ordering, and Kueue for resource admission \cite{Tejedor2026-pg}. The work does not address contention, co-allocation, 
or attribute-based selection of the quantum backend itself.

\smallskip
\noindent{\bf Quantum Workload Management with Fluxion} 
Relative to this Kubernetes prior work, our contribution is the use of a graph-based scheduler from the Flux Framework (Fluxion, via Fluence \cite{fluence, 9652595,anon-converged-scalability-2022}) to model a remote quantum backend as
a first-class, attributed, arbitrated resource. With support for native gang co-scheduling in Kubernetes 1.36.1 \cite{gang-scheduling}, our approach allows for modeling of different vendors with selection of attributes in the scheduler resource graph, and coordination with classical compute resources. We make the following contributions:

\begin{itemize}
 \item{Characterization of the \textit{two-queue problem}}
 \item{Graph models of hybrid classical and quantum systems}
 \item{Quantum scheduling designs based on application needs}
 \item{Kubernetes scheduler plugin for custom resources}
 \item{Reproduction of gang scheduling experiments}
 \item{kubectl plugin for user-space match policies}
 \item{Experiments to demonstrate two-queue problem}
\end{itemize}

In Section \ref{sec:methods} we discuss our approach to model quantum resources in the Fluxion resource graph, and monitor queue depth to dispatch workers only when the job is ready for execution. We present experiments that measure latency from using gang scheduling alone as compared to Fluence, manual and policy-based selection of quantum vendors based on queue depth and price, and dealing with job contention. In Section \ref{sec:results} we share results of our work, and finish with discussion (Section \ref{sec:discussion}) and conclusions (Section \ref{sec:conclusion}).

%% file: sections/2-methods.tex
\section{Methods}
\label{sec:methods}

\label{sec:overview}
\subsection{Overview}
We present a design for a scheduling layer in Kubernetes that allows for hybrid coscheduling of quantum devices with classical compute nodes. We argue that the scheduler should treat the quantum backend as a first-class resource and defer the classical workers until the quantum task is near ready, rather than admitting them to idle.  Our approach enables attribute-based selection of quantum backends, or more generally, custom resources that are known to the scheduler. Unlike other approaches, our solution restricts credential access to the owning user. 
We start with Kubernetes as a resource manager due to ease of development, and an analogous design will be extended to the Flux Framework for traditional \gls{hpc}.

\begin{figure}[ht!]
  \centering
  \includegraphics[width=\columnwidth]{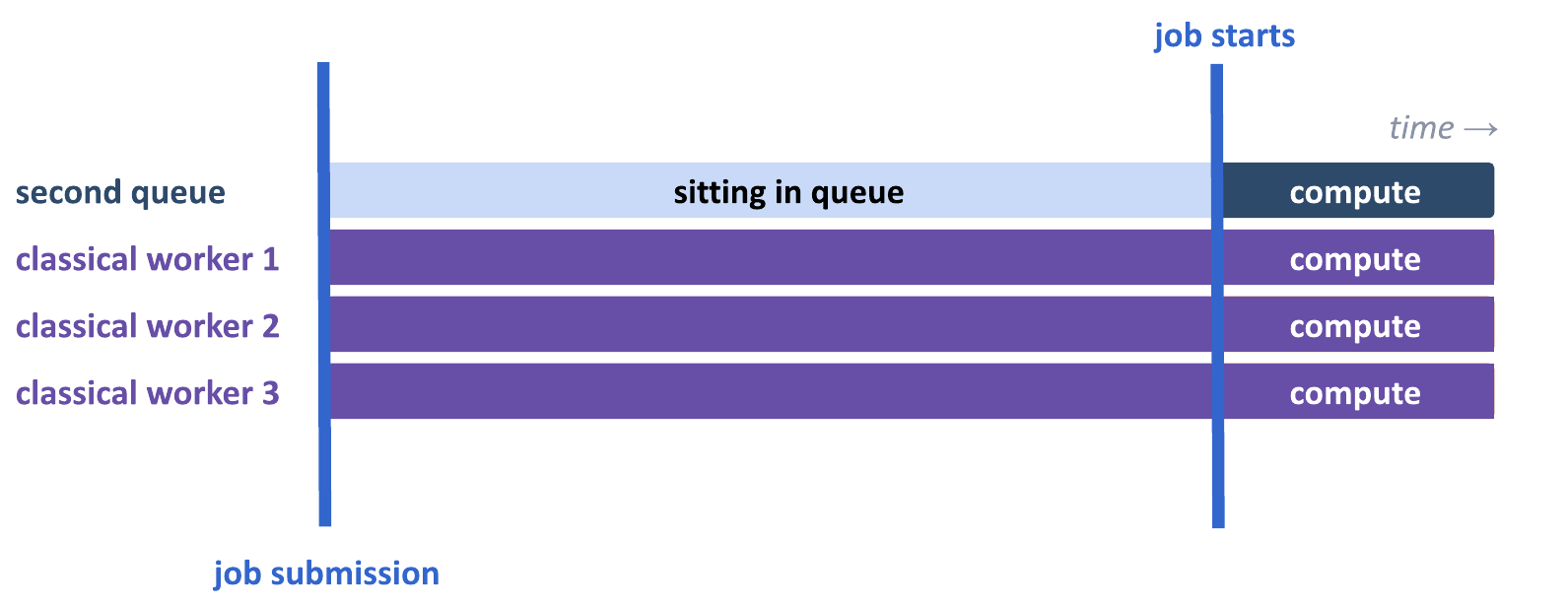}
  \caption{\textbf{The two queue problem.} \normalfont  Scheduling one workload manager's resources against work waiting in a separate queue that is neither controlled nor completely observable by the workload manager. Given a dependency structure between the two systems, resources are required to wait in absence of intelligent co-scheduling.
}
  \label{fig:two-queue-problem}
\end{figure}

\label{sec:two-queue-problem}
\medskip
\noindent{\bf Two Queue Problem} 
The two-queue problem (Figure \ref{fig:two-queue-problem}) refers to the need to schedule one workload manager's resources with work waiting in a separate queue that is neither controlled nor completely observable by the workload manager. In the domain of quantum computing, the second queue can be a vendor queue, or remote queue at a different institution. Without a solution, there is no way to predict the waiting time for resources in the second queue, and the result is pathological behavior of allocating and idling resources. With a solution, the workload manager can schedule and orchestrate successful co-scheduling synchronous in time. This problem has been described for analogous setups \cite{badts2026examiningqrmiunifiedinterface}.

\label{sec:kubernetes}
\smallskip
\noindent{\bf Kubernetes} 
 We choose Kubernetes as our workload management environment for this initial prototype due to the ease of development with the scheduler Fluxion, which can be used as an independent component of the Flux Framework, and exposed via the Go \gls{sdk} \cite{ahn-2014}. We developed a new variant of our 2021 project Fluence \cite{9652595} with the intention of supporting native Gang Scheduling and custom resources. With Kubernetes we can use a declarative model that is not typical of \gls{hpc} workload managers: modeling the custom resources in the Fluxion resource graph, allowing for a Pod specification to request \gls{qpu} devices, and using Kubernetes scheduling gates to gate workers until the quantum work is ready to run.

\label{sec:graph-modeling}
\smallskip
\noindent{\bf Graph Modeling} 
Fluxion models cluster resources as a graph, where the top level is a root (cluster) and children progress as racks, nodes, sockets, and then cores, GPUs, or similar hardware. A request for work is represented in a job specification, which visually shows a shape and count of resources to be searched for in the graph. For example, a user might be interested in two sets (each set called a ``slot'') of 32 cores that, for each set, are under a common node. A different request might ask for two slots with 32 cores under any number of nodes. Finding the number of slots to satisfy the request comes down to a graph traversal over time based on a match policy. For example, a ``first'' policy will return the first match found.

\begin{figure*}[ht]
  \centering
  \includegraphics[width=1.0\textwidth]{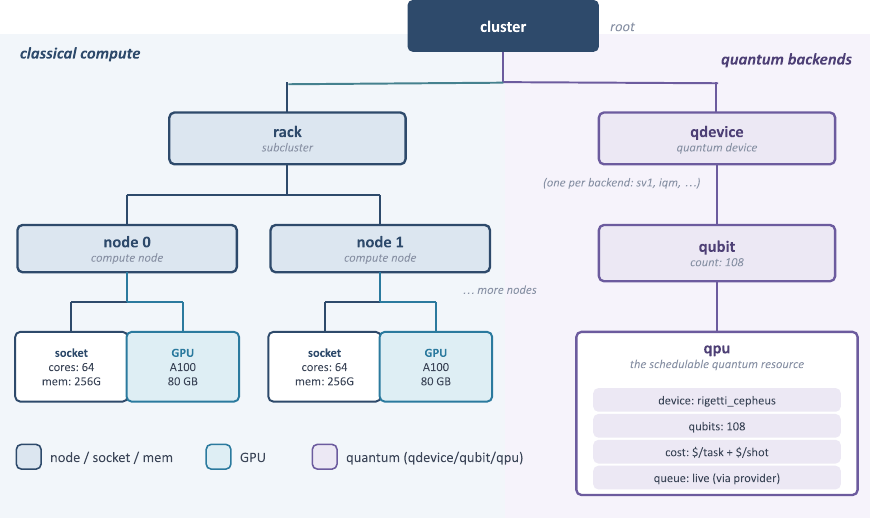}
  \vspace{-20pt}
  \caption{\normalfont \textbf{Hybrid modeling of classical and quantum resources} in the Fluxion resource graph. A quantum device is modeled at the same level as a rack, and job requests for work correspond to a traversal in the graph. Constraints are used to limit search to a resource class.}
  \label{fig:fluence-design}
\end{figure*}

The graph modeling is powerful because any hierarchy of resources can be represented in the graph, and queried over with a job specification. For our quantum work, recognizing that quantum devices abstractly lived under the same cluster, but alongside classical resources, we place top level quantum devices, each associated with a vendor, at the same level of a rack (Figure \ref{fig:fluence-design}). While a search can be done across the graph, we opt for an approach that does a separate query for each of quantum and classical resources using constraints \cite{flux-constraints}. Each query immediately prunes away the other resource type to reduce the search space. Both queries are done back-to-back by the scheduler, and only allowed to move forward to schedule work if both resource types can be allocated. Constraints can also be used to select resources based on arbitrary properties like vendor, quantum device, or quantum class.


\begin{figure*}[ht]
  \centering
  \includegraphics[width=1.2\textwidth]{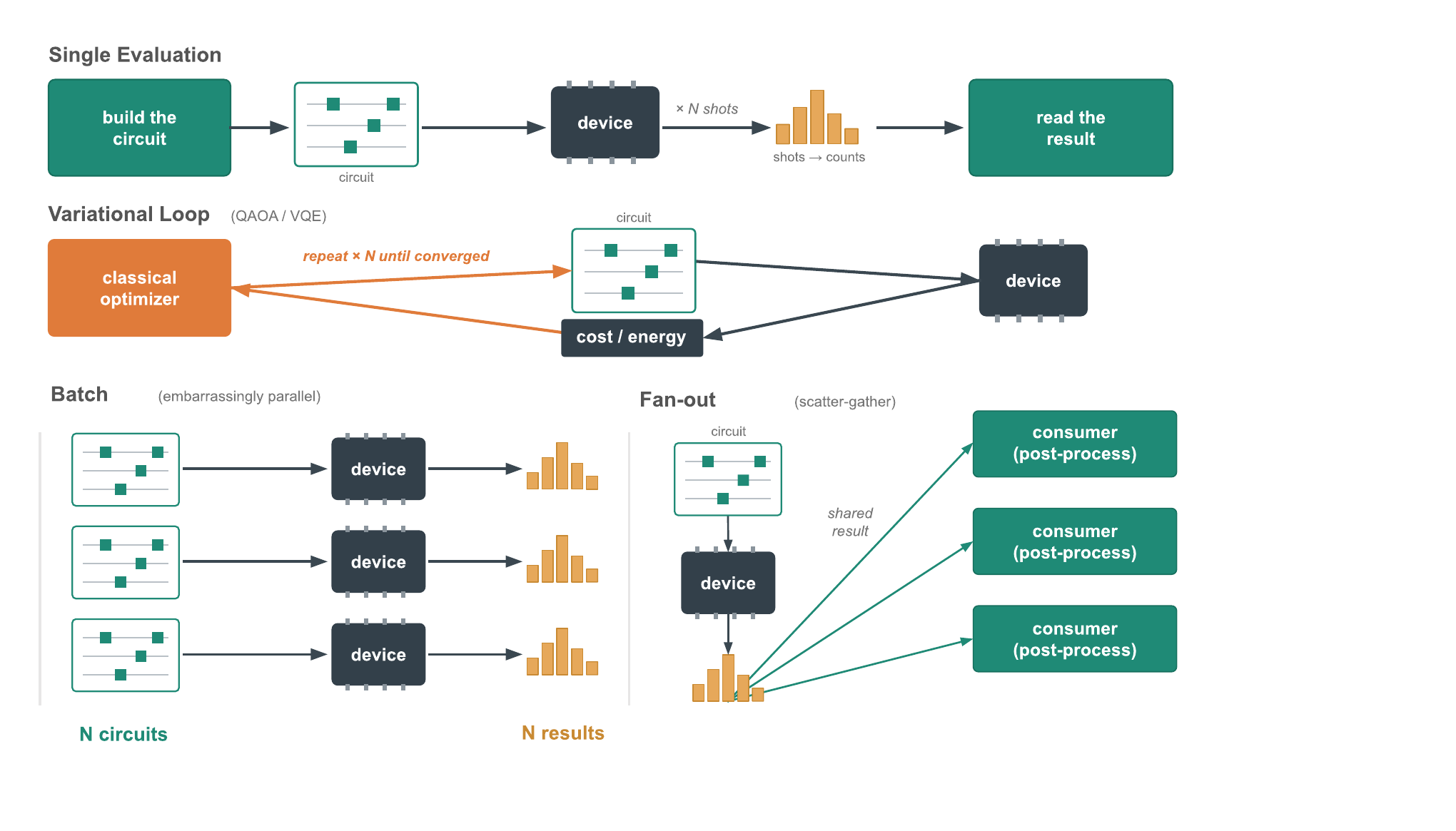}
  \vspace{-50pt}
  \caption{\normalfont \textbf{Hybrid quantum and classical application designs}. The single evaluation (top), when a circuit is run and post-processed, forms a simple, single execution  unit that is the basis for more complex designs. The variational design (second row) maps this unit into a cycle, receiving back results, and adjusting parameters to run again until convergence. The batch pattern (bottom left) runs many single evaluations in parallel to generate the equivalent number of results, and a fan-out pattern (bottom right) provides one quantum result for many classical consumers. These designs pair with the scale, frequency, and coordination of execution, form the basis for hybrid quantum scheduler designs. 
}
  \label{fig:app-designs}
\end{figure*}

\label{sec:designs-for-quantum-apps}
\smallskip
\noindent{\bf Quantum Application Designs} 
The design of a scheduler must be suited to the applications it serves. Many quantum algorithms are hybrid, meaning a list of operations, a circuit, is assembled and sent to a quantum device. Each time that a device runs and returns probabilistic output in the form of a measurement, a shot, the classical resources are typically responsible for post-processing. If necessary, more calls are made. The design differences between applications come down to the pattern and frequency of task/circuit submissions between the two environments. A depiction of common designs is shown in Figure \ref{fig:app-designs}.

The simplest application pattern is a \emph{single evaluation}, when one circuit is built, run, and evaluated. There is one call to the device, and this design forms the unit that constitutes more complex designs. A more common pattern is a \emph{variational loop} that repeats this pattern in a cycle, meaning that between quantum tasks, a classical computer changes parameters to resubmit. The process continues until some convergence or goal is reached. This pattern is used by QAOA and VQE, and is the most challenging for two queues to handle given the unpredictability of queue depth and running times. In a batch, N circuits are independently generated, possibly with different parameterizations of a problem, and submitted in an embarrassingly parallel style. Finally, for a fan out or scatter gather design, one quantum result feeds many classical consumers. These application designs pair with the scale, frequency, and coordination of execution across environments to inform quantum scheduler designs.

\label{sec:designs-for-quantum-scheduling}
\smallskip
\noindent{\bf Quantum Scheduling Designs} 
The scheduler designs best suited for quantum application designs depend on the level of control of the quantum resources. There are two models for hybrid interaction, and they come down to having control of the quantum devices. In a \emph{remote} or vendored model, the scheduler is modeling and submitting to remote \gls{api} resources that are not directly controlled. As an example, using IBM Quantum Cloud or Amazon Braket introduces a second queue. A graph scheduler can model resource counts, quotas, and topology, however an immediate ``allocation'' of paired quantum and classical resources cannot guarantee that the classical resources will run at the same time as the quantum. More realistically, the contention of the quantum devices and tendency to have queues will have classical resources like GPUs idling for an unspecified amount of time. We have introduced this problem of coordination as the \emph{two-queue problem}, where a resource manager is modeling resources for jobs that are dispatched to a second queue not under direct control. This problem is distinct from the case when an allocation of resources at a specific time guarantees exclusive and immediate access. Remote devices could reliably guarantee execution times if a reservation can be made, or immediate execution otherwise guaranteed.

The details of each vendor vary subtly. With Amazon Braket, each task submission is submitted to a shared queue that is subject to waiting time. Customers can only get priority treatment if they use the Amazon Braket Hybrid queue, which requires using EC2 instances as classical and could not be extended to a local \gls{hpc} scheduler classical resources or an external Kubernetes cluster. Reservations by the hour are also possible, with costs ranging from \$2,500 to \$7,000 an hour \cite{braket-prices}. Thus, while there are established means to orchestrate workloads with a priority queue, the approaches are expensive or require using cloud classical resources. Amazon Braket automatically saves quantum results to S3 storage, allowing for applications in Kubernetes that do not have a shared filesystem to discover them based on sharing a task identifier. The \gls{qrmi} interface provides an abstraction to vendors such as IBM Quantum Cloud. With IBM Quantum Cloud, creation of a session will give subsequent jobs in the queue priority under the constraints of a time to live (TTL) \cite{ttl}.

\label{sec:fluence-design}
\smallskip
\noindent{\bf Fluence Design} 
We present \emph{Fluence}, a Kubernetes custom scheduler plugin that models custom resources such as quantum devices directly in its resource graph, and comes with support for gang scheduling and intelligent design to coordinate custom resources with classical work. The value of Fluence is no longer to solely create gangs, but rather serve as a scheduler plugin that is optimized to customize Pod groups on the fly based on requested resources and application design patterns. Following our design from 2021 \cite{9652595}, we recreated Fluence with several improvements. First, we use native gang scheduling afforded in Kubernetes 1.36.1 enabled with feature gates \cite{gang-scheduling}. Second, we removed the design that deployed Fluxion as a sidecar service to the scheduler plugin, removing unnecessary communication between two pods. Finally, we add support for custom resources to automatically be added to the resource graph and available for scheduling against. 

Any Kubernetes scheduler plugin must define one or more extension points to influence the scheduling cycle that correspond to states of a Pod in the single queue \cite{scheduling-framework}. Example extension points include filtering to narrow the set of feasible nodes, scoring to rank nodes, and binding to allocate the Pod to a chosen node. A custom scheduler plugin can also use a handle that exposes an abstraction called an informer that serves to deliver events to inform scheduling. For example, an informer can receive Pod deletion notifications to know when a Pod is terminated. Fluence implements the extension points of \emph{PreFilter}, \emph{Filter}, \emph{PostFilter}, \emph{Reserve}, \emph{Unreserve}, and \emph{PreBind}. 

A request for work that is tagged with \emph{schedulerName: fluence} proceeds as follows. The Pod prepares to enter the scheduling queue tagged for Fluence. Before reaching the queue, a mutating admission webhook owned by Fluence checks details of the group, and allows several handlers (e.g., \emph{quantum} or \emph{gang}) to look over resource requests and annotations to decide if custom handling logic is needed. The webhook can mutate a Pod or PodSpec accordingly to customize the gang behavior, for example injecting annotations that can be populated after scheduling or adding provider-specific sidecar containers. The pods for the group then enter the scheduling queue tagged for Fluence and proceed through the plugin's extension points. In \emph{PreFilter}, the first Pod of a group to be processed triggers a single Fluxion match for the entire group. The scheduler graph that includes custom resources is populated by a cluster \emph{ConfigMap}, which allows an admin to define roots to insert trees of children nodes that have classes and properties to represent schedulable resources. Fluence is agnostic to the handler logic, and takes an approach that generates a job specification ``jobspec'' and does a graph query (a \emph{match\_allocate} \cite{rfc-27}) for each resource type, treating them as a pair that must be scheduled together, all or nothing. Conceptually, the job specification represents a request for a count of a resource shape to be searched for in the graph. A successful match can find the correct count of the shape, and mark the vertices as utilized while the job is running. In the future when Flux Framework has full support for subsystems, one or more custom device types can be done as a single match to the graph \cite{flux-subsystem-discussion}. 

\label{sec:fluence-handlers}
\smallskip
\noindent{\bf Webhook Handler Interface} 
The Fluence webhook has a handler interface that can define custom logic for one or more handler types. The default handler is for a gang or group of pods. For example, for an abstraction that requests Fluence as the scheduler without any custom resources, a group of size N is created. Single pods are treated as groups of one. For this work we focus on custom resources in the form of remote quantum devices, and extend our example to walk through the case of a Job requesting quantum resources. The Fluence webhook has a handler interface for which quantum is defined as a type. For a Job of size N, the quantum handler, upon detecting a request for quantum resources, creates two groups of 1 and N-1, where the single Pod group (typically index 0 for the Job) acts as the producer, and the remainder N-1 pods are workers or consumers.  The webhook injects annotations that will be updated after scheduling and populate environment variables for the quantum application to discover. An example application for Braket might be the Amazon Resource Name (ARN) that determines the remote quantum device. A quantum-specific backend sidecar (e.g., Braket or \gls{qrmi}) is also added to the producer Pod that will orchestrate ungating. The worker (consumer) Pods are gated until the remote quantum task has reached the top of the queue, determined by the producer sidecar polling it. This design means that Fluxion will not schedule the classical gang until the Pods are ungated. The pods for the group then enter the scheduling queue tagged for Fluence and proceed through the plugin's previously described extension points, targeting quantum devices as custom resources that are defined in the Fluence resources \emph{ConfigMap}. To ensure that ungated Pods are scheduled promptly, they are given a higher priority. In the case of a filled cluster, lesser priority Pods will be evicted to make room. This same design could be extended to a traditional workload manager given a pending job with priority that is ready to run paired with a service to act as the quantum monitor.

In \emph{Filter}, Fluence constrains the Pod to the node or nodes its allocation selected. A device-only allocation pins no node, because the backend is a remote \gls{api} reachable from any node, so Filter imposes no constraint. In \emph{Reserve}, the selected backend name and attributes are annotated onto the PodGroup, allowing the webhook's environment contract to resolve them into the environment. The corresponding \emph{Unreserve} is a deliberate no-op, since a stale annotation from a rejected reservation is harmless and is overwritten on the next attempt. Finally, in \emph{PreBind} Fluence records the Fluxion job identifier onto the same PodGroup, allowing for a cancel call to Fluxion when the Pod completes or terminates, freeing resources in the graph. The scheduling framework does not provide an extension point for Pod termination, and so Fluence uses informers to be notified of the events. 

The ability of Fluence to handle different quantum application patterns depends on the backend. For example, in that \gls{qrmi} can create an initial session with a waiting period only for the initial task, all patterns (single evaluation, variational loop, batch, and fan-out) are supported in that subsequent requests are within the batched group that has priority, and the worker pods ungate after the first task reaches the first position in the queue. Amazon Braket can only well-handle the single evaluation and fan-out pattern, as any second call that does not have priority will be subject to potentially a full waiting period after Pods are already running. 



\label{sec:experiments}
\subsection{Experiments}

\label{sec:gang-selection}
\noindent{\bf Gang Selection} 
To validate Fluence as a gang scheduler and reproduce the result in our first paper \cite{9652595} we aimed to run an experiment that would stress-test gang scheduling, deploying three \gls{hpc} applications (AMG, LAMMPS, and QMCPACK) across three sizes (N=1,2,4) on a small cluster (N=4 nodes) to create contention. Each application Pod added resource requests to occupy the entire node using a Flux Operator MiniCluster \cite{anon-flux-operator-2024} that has recent support to specify the scheduler name to the Job PodTemplate. We measure contention, and specifically the time that partial gangs are scheduled and not able to run until smaller gangs clean up. A successful refactored design of Fluence would reproduce the first result to show longer end-to-end time of the first Pod of a gang coming up to the entire gang finishing, where the time of partial pods waiting for resources represents poorly utilized resources. We tested up to 36 gangs per scheduler across sizes, where each set of gangs is randomly shuffled before submission at the same time, and found that 9 gangs was the only size that would reliably create contention without the queue locking due to partially scheduled gangs for the default scheduler. We chose to run two iterations using the maximum of 9 groups to avoid this blocking and still get an assessment of extra overhead for partially running gangs that would eventually be unblocked to run. 

\label{sec:policy-selection}
\smallskip
\noindent{\bf Policy Selection} 
Across application designs, it is often desired to choose the best device for a job, where ``best'' is subjective, but might be the quantum resource that minimizes queue wait time or cost. As the first requires a user credential to query a provider backend, we implemented a \emph{kubectl} plugin for Fluence \cite{kubectlfluence}, and reiterate the ability of Fluxion to allow for query and filter based on properties in the resource graph. We implemented this as a local plugin not due to technical constraints, but the desire to keep \gls{api} calls that require credentials local. For this experiment, we will do 10 calls of using Fluence with and without the kubectl plugin, asking to minimize cost or queue depth, and demonstrate a lower mean cost and waiting time when using the plugin as opposed to choosing a backend at random. The plugin works by way of using a ConfigMap owned by Fluence that populates the resource graph with the \gls{api} custom resources. Any node property that is added to the graph is available as a feature to minimize, maximize, or otherwise query. This backend-selection mechanism is dynamic, and enables custom policies per-job, a feature not available in existing scheduler plugins for \gls{qrmi}, or even Fluxion natively. Native Fluxion allows for exact string matching, but not evaluation of operands against numeric values.

\label{sec:quantum-selection}
\smallskip
\noindent{\bf Quantum Selection} 
For our main experiment, we will demonstrate Fluence acting to run quantum gangs of workers, where each gang will submit work to one of 6 quantum queues: Amazon Braket simulators \emph{sv1} (state-vector), \emph{dm1} (density-matrix), \emph{tn1} (tensor-network), and real quantum providers \emph{Rigetti Cepheus}, \emph{IQM Emerald}, and \emph{IQM Garnet}. The first Pod in the set acts as the producer, building the QAOA max-cut circuit, submitting the task, and publishing the task identifier. The other N-1 pods are consumers in that they do not submit tasks, but obtain the producer's task identifier, fetch the shared result, and process their shot partition. This application design is the fan-out pattern, and given a backend like \gls{qrmi} that supports sessions with priority, it would also work well for repeated calls done in batch and variational loops. Under Fluence with quantum custom resources, the producer Pod requests the \emph{fluxion.flux-framework.org/qpu} resource, triggering Fluence to match paired quantum and classical resources. The producer is scheduled first, and the classical gang only after the producer ungates them. 

We will demonstrate that an approach using Fluence where a single producer submits the work and the N-1 worker pods are gated until the task reaches the top of the queue reduces worker idle time to more effectively use classical resources. Under the default scheduler, an analogous gang is co-scheduled with a native Kubernetes \texttt{scheduling.k8s.io/v1alpha2} PodGroup. All N pods will start together, all-or-nothing. We will mirror the producer/consumer roles from the Job completion index, and add the device name and task identifiers by hand, ensuring that despite different scheduling, the application runs equivalently. As nothing gates the consumers, they will run idle on real nodes for the entire wait.  With Fluence, we will demonstrate that there is essentially no idle time, as the workers are brought up synchronously when the quantum task is ready. Without Fluence, the entire gang will come up and be subject to occupying the classical resources during the same waiting period. For these experiments, we chose three simulators and three real quantum queues that were reasonable under cost constraints. Since Fluence does not model actual, controlled devices in the Kubernetes cluster, the shape of the resource graph would add value to represent quotas and features afforded by the different simulators or real quantum systems. 
We will calculate a total consumer idle time in seconds as the sum of time over consumers that are running before they successfully retrieve a task identifier associated with a result, and run experiment sweeps across group counts of 2, 4, and 8 with 10 repeats for each of Fluence and the default scheduler.

%% file: sections/3-results.tex
\section{Results}
\label{sec:results}

\label{sec:gang-selection-results}
\begin{figure*}[ht]
    \centering
    \includegraphics[width=\textwidth]{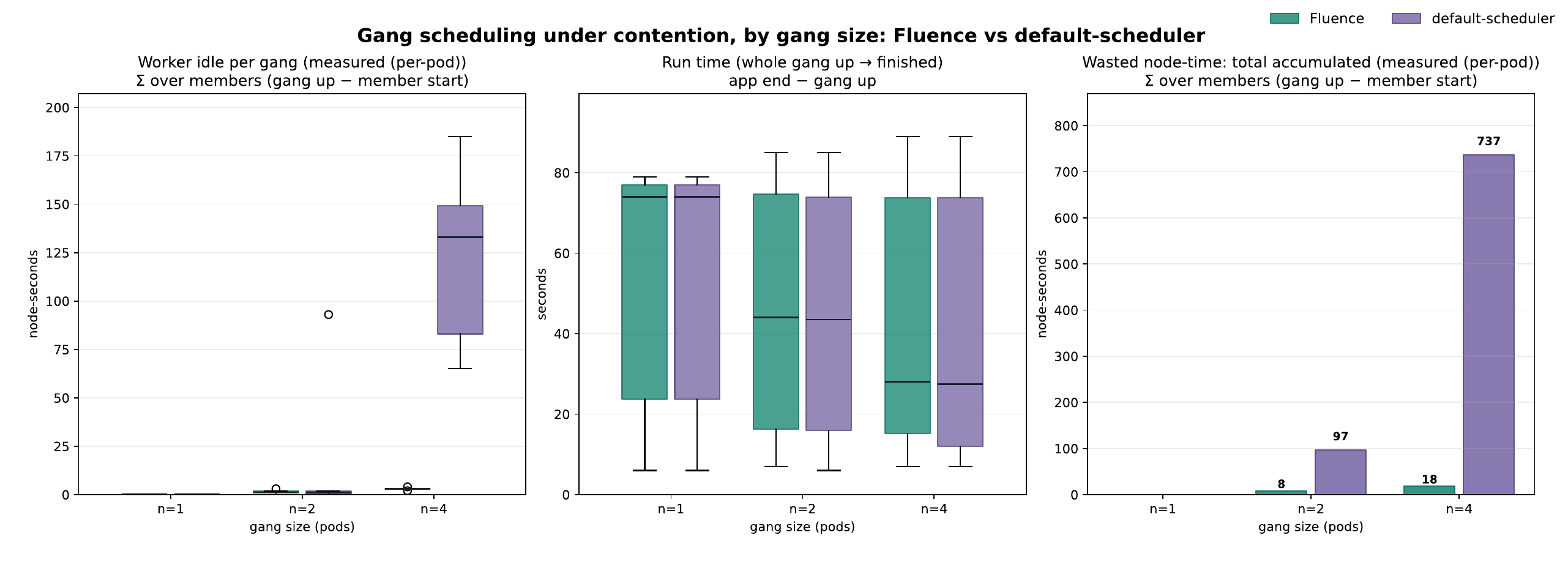}
  \caption{\textbf{Gang scheduling for 18 gangs (9 gangs × 2 iterations) of varying sizes across 3 applications} for each of Fluence and the default-scheduler. \normalfont Worker idle time increases with size (left panel), and accumulated node time (right panel) represents poor resource utilization across workers in the cluster. There are no differences in application running times once all members of gangs are ready (middle panel).}
  \label{fig:gang-selection}
\end{figure*}

\smallskip
\noindent{\bf Gang Selection} 
We ran an experiment to stress-test gang scheduling by deploying three \gls{hpc} applications (AMG, LAMMPS, and QMCPACK) across three sizes (N=1,2,4) on a small cluster (N=4 nodes) to create contention. Despite choosing a reasonable size (N=9 gangs) one default-scheduler attempt deadlocked on partially scheduled gangs and produced no results. Reported data include two completed iterations of 9 gangs per scheduler.  Results are shown in Figure \ref{fig:gang-selection}. We demonstrate that worker idle time, when a partial gang is running and waiting for quorum, increases with size, with Fluence keeping the idle time close to 0. Smaller gangs of size 1 and 2 served primarily as filler, and we did not observe them to be commonly waiting, with the exception of one gang for the default scheduler that waited for $\sim$93 seconds. Single pods for both schedulers by design have no worker idle time, as the single Pod can run independently. Separation between behavior appears at size N=4, where the default scheduler's worker idle time jumps to a median around 133 node-seconds. This behavior results from partially scheduled gangs that have workers that must wait for smaller gangs to finish before coming up. Fluence's all-or-nothing placement removes that wait, so idle stays flat regardless of size. The summed, total accumulated wasted node time is shown in the third panel, with the largest difference at size N=4, where the default scheduler has workers waiting idly for 737 seconds as compared to Fluence's 18. This small amount of time represents the time the Init Containers, which set up the Flux install, take to run between Pending and Running. Importantly, the middle panel shows no differences in the running time of the applications once all gang members are ready.

\begin{figure}[ht]
    \centering
    \includegraphics[width=\columnwidth]{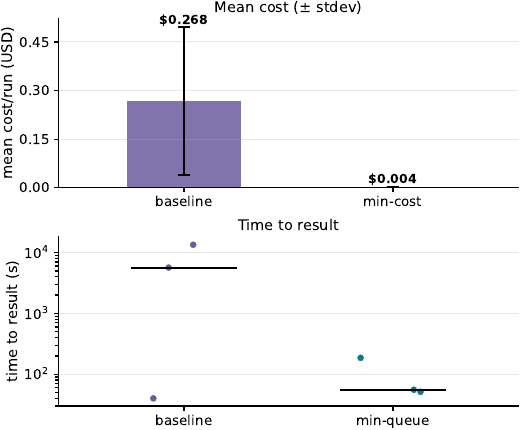}
  \caption{\textbf{Policy aware selection}. \normalfont Requesting Fluence to select a minimum cost backend clusters all points around a near-zero cost (top panel). Asking Fluence to select minimize queue depth leads to a faster end-to-end time to complete the work (bottom panel). }
  \label{fig:cost-selection}
\end{figure}

\label{sec:policy-selection-results}
\smallskip
\noindent{\bf Policy Selection} 
We ran experiments to assess the impact of a Fluence policy to minimize cost or queue depth on the outcome of those variables (Figure \ref{fig:cost-selection}). Requesting Fluence to perform selection to minimize cost results in a lower mean cost as compared to random selection. Asking Fluence to select based on minimum queue depth, while queue depth cannot be a complete indicator of estimated time to arrival of a quantum job, does lead to a faster end-to-end time and is a reasonable strategy to more efficiently complete work when selecting between comparable backends.

Using three QPUs and three simulators, the random baseline included more expensive devices like \emph{iqm\_garnet} and \emph{iqm\_emerald}, and the min-cost policy pinned \emph{dm1} on all runs, resulting in a cost reduction of approximately 72x and eliminating variation. The selection of devices includes real quantum devices that had no queue depth at the time of a subset of runs. Notably, queue depths are out of user control and largely unpredictable.

\begin{figure*}[h]
    \centering
    \includegraphics[width=\textwidth]{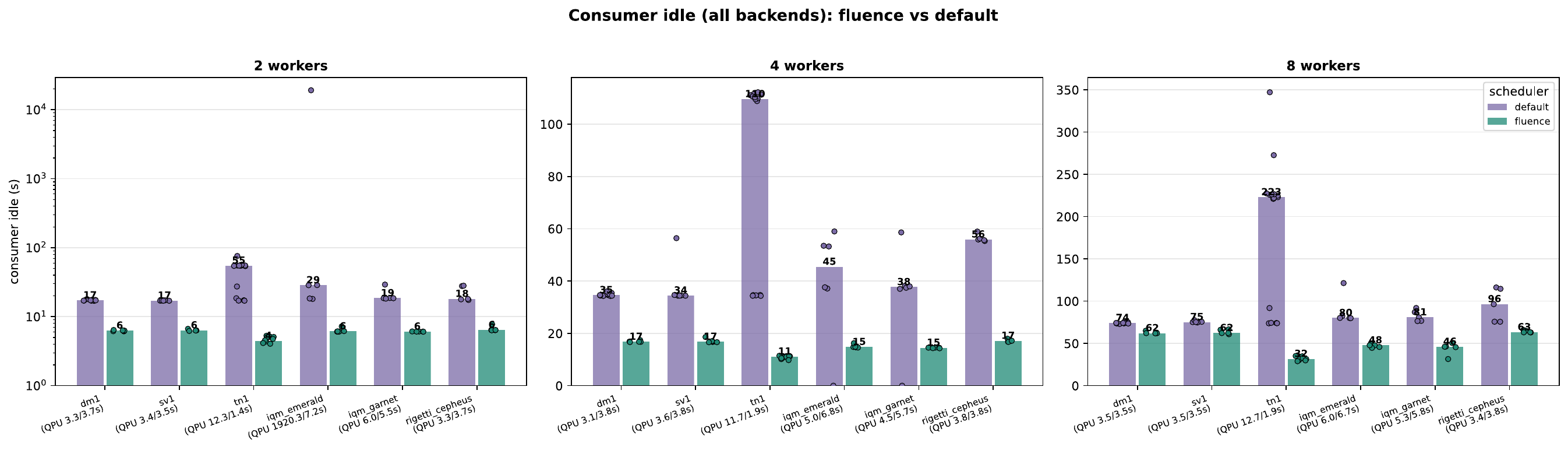}
  \caption{Worker idle time across quantum simulators or devices, compared to number of workers. As expected, the idle time increases with the size of the job.}
  \label{fig:worker-idle-time}
\end{figure*}

\label{sec:quantum-selection-results}
\smallskip
\noindent{\bf Quantum Selection} 
The quantum selection experiment results are shown in Figure \ref{fig:worker-idle-time}. Each run is a gang of N pods with a producer submitting the Braket task and N-1 consumers waiting on it. Across three real QPU backends and three simulators we observe that Fluence's idle time is lower for every backend and group size. The gap widens with both consumer count and per-task service time. At N=8 consumers the default scheduler's median idle is 74 s (\emph{dm1}), 75 s (\emph{sv1}), and 223 s on \emph{tn1} as compared to Fluence's 62, 62, and 32 s, respectively. \emph{tn1} is a tensor-network simulator with a longer per-task service time than the state-vector and density-matrix simulators. Using Fluence, worker Pods are gated until the quantum task is ready to run, and under the default scheduler, all nodes wait, demonstrating excess worker idle time as a result of the two-queue problem. For the gangs of size 2, the plot time (y axis) is shown with a logarithmic scale due to a single 2.66 hour device queue on \emph{iqm\_emerald} resulting in 19,160 node seconds of waiting. While we cannot report the frequency of this kind of wait, it is not uncommon and is a real-world example of why co-scheduling strategies are needed to schedule and dispatch resources. We suspect that the waiting time without such a mechanism is potentially unbounded.

%% file: sections/4-discussion.tex
\section{Discussion}
\label{sec:discussion}

We have presented Fluence, a custom scheduler plugin for Kubernetes that can handle gang scheduling and custom resources such as remote quantum devices, demonstrating an ability to intelligently schedule gangs of pods for quantum applications that coordinate classical resource scheduling with remote quantum \gls{api}s. Our work is important because no current existing implementations for quantum \gls{api} interaction consider this co-scheduling problem to coordinate two queues, and model the quantum resources as first-class citizens in the resource graph.  There are several points for discussion.

\label{sec:vendor-design-requests}
\smallskip
\noindent{\bf Vendor Design Requests} 
The designs we have described are largely developed to get around the constraints presented by a vendor API. While it is advantageous to query for a queue depth, an event-driven architecture would efficiently allow for a client to receive notice of a quantum task that is ready to run. The language that is used for \gls{api} endpoints can also be misleading. For example, the \emph{acquire} function to start an IBM Quantum Cloud session suggests a successful execution to retrieve a scoped set of resources for work. In practice, the behavior of the call can vary based on the vendor backend. The first task can still be  subject to the same queue time. A design that would work well to support co-scheduling would allow for the acquire function to return only when the client has moved up in the queue and is ready to serve tasks with higher priority. The design would allow for placing a foot in the queue without submission of actual work by the scout or producer that starts the session. Amazon Braket does not go so far as to allow remote calls to an equivalent endpoint to get placed on a priority queue, and only allows for a hybrid execution that requires using hosted EC2 instances. From a consumer perspective, the requirement comes across as an attempt to force customers to remain locked into cloud resources. 
The alternative on Amazon Braket, the creation of reservations, is limited to one hour blocks that can cost in the thousands of dollars, a price that may not be feasible for all groups. More control and ability to get a priority queue while using local resources would hugely support the \gls{hpc} community.

\label{sec:homogenous-gangs}
\smallskip
\noindent{\bf Homogeneous Gangs} 
For remote vendor \gls{api}s, we do not see a strong design use-case for homogeneous gangs, where each gang member schedules independent quantum resources. The reason is because this design would deploy N pods at the same time, each with a sidecar to monitor the queue, and incurring the cost of all N pods waiting. A more plausible design would be to have a producer submit a batch of work, and then disperse the request identifiers to the worker pods, ungating when the batch is at the first position in the queue. If a user did want to deploy N pods that each submit an independent task to the queue, then Fluence is not needed to do the scheduling, because there is no scheduling coordination to be handled. The user accepts the idle time of each worker and has application logic dictate independent submissions. Equivalently, the user can pay for an expensive reservation of a quantum device and not be subject to any queue waiting time.

\label{sec:workflows}
\smallskip
\noindent{\bf Workflow Integration} 
Much of the plugin logic \cite{Bacher2025-hi} is handling what we see as being under the jurisdiction of the user application, or better, a workflow tool. To make the distinction, a workflow tool is responsible for defining units of work, and submitting a unit (job) to a workload manager that requests the quantity and type of resource needed for a step. A workflow tool would ask a workload manager for a quantum device with specific properties, and the workload manager would decide on the exact device, perhaps following a policy related to time available or cost, and coordinate the scheduling of classical and quantum resources. Validating that a specific set of environment variables are exported, a task that is likely to be re-used across quantum workflow pipelines, is an ideal task for a workflow tool. In the case that quantum applications will perform repeated steps of similar work that primarily differ in input parameters, a workflow tool would be ideal to define the set, allowing for definition of ``best practices'' by experts directly into the tool. 

Workflow tools are also oriented to allow for specifying resources that are needed for specific units of work. For example, a quantum application that will run a variational loop likely will want a scheduler with support to request a priority queue. The discussion of these interfaces (e.g., \gls{qrmi} and Braket and specific applications) needs to be continued, as it is easy to confuse workflow tools with workload managers, and it should not be the case that a workload manager is performing tasks that should be done by the workflow tool.

\subsection{Limitations}
We were not able to dispatch a larger set of gangs for our group experiment due to partially scheduled jobs clogging the queue for the default case, and instead chose to do two iterations of a smaller set of gangs.  We are limited to the design of our Fluxion interactions based on upstream features.

For our quantum selection experiments, we note that the 2-3 hour waiting time by chance hit the default scheduler run, and could have easily been allocated to Fluence. However, we also note that Fluence would have had the equivalent ideal behavior in trivial worker time, because the workers would not have been dispatched.
More work is needed to understand patterns of vendor queue availability and queue depth.

\subsection{Future Work} 
\label{sec:future-work}
Fluence represents a remote-only design for which we model resources that are not under our control, and use scheduling mechanisms to make a best effort to have work start synchronously. A second use case is the one for which we can more directly control, or have a guarantee about exclusive access to a specific set of quantum devices. As an example, local quantum machines that are controlled at an institution can be scheduled akin to compute nodes, ensuring that a block in the workload manager schedule corresponds to known, specific devices or channels, and during that block, other users are not able to access them. The case is analogous to using cloud versus local high performance computing in that provisioning and getting access to cloud resources is often subject to a similar waiting period, a problem that is compounded by resource contention \cite{anon-cloud-usability-2025}. Notably, a device does not need to be directly controllable if it is reliably reservable. For example, a future reservation of quantum from either a vendor or collaborator institution can equivalently be blocked in the resource graph, and started alongside classical resources. The strength in our design is that the modeling itself -- the representation of quantum devices in a graph -- is the same strategy to take either way. The differences in execution come down to how the scheduler is deployed. In Kubernetes that delivers a single queue of pods, we use a custom scheduler plugin and gating to hold off on scheduling until the quantum resources are ready.  On \gls{hpc} we are working on similar design strategies, still using the Fluxion scheduler, but with Flux Core plugins to support the design. 
An interesting area of work will also be thinking through how a Fluxion-based custom scheduler plugin could support a Reservation abstraction in Kubernetes.

We chose to implement policy selection in the user space, allowing the user to use local credentials to query for queue depths and costs, and mutating a Pod or request for work with a constraint to specific quantum providers. An improved approach would be to have a Policy interface implemented alongside Fluence, however this introduces the challenge of requiring some credential to get access to the equivalent \gls{api}s. Any approach that places user shared credentials for submitting work in the scheduler space we view as a security risk, however it could be done to use a scoped credential that would only be able to query for queue depths and retrieve costs. Using this approach, given that a user selects a quantum device type, the specific quantum vendor and machine could be selected to minimize costs and/or queue depth. Our policy experiments can and should be improved to not exclusively select one resource, for example, selecting devices under a specific cost and then randomly selecting from that subset.

Finally, our implementation in Fluxion that does two all-or-nothing queries to schedule each custom resource is functional and reliable, however an improvement will come when Fluxion is able to support subsystems, meaning that different resource types can be represented and queried under a common root.  While we used Amazon Braket for our experiments, the same design will apply to \gls{qrmi} that first does an acquire call before task submission. Our inability to run experiments here only came down to not having a working \gls{api} token with a paid account at the time of execution. With \gls{qrmi}, and specifically using IBM Quantum Cloud, there will be an improved orchestration and support for workloads that require more than one call (the batch or variational loop pattern) as after the first call and queue waiting time, subsequent jobs are given priority \cite{ibm-quantum-cloud}. A preferential priority queue is only available on Amazon Braket as a hybrid quantum submission, which requires using EC2 instances.

\section{Conclusion}
\label{sec:conclusion}
We have presented Fluence, a Kubernetes scheduler plugin backed by the Fluxion graph-based scheduler, which models remote quantum backends as first-class, attributed vertices in the resource graph and co-schedules them with classical compute. We re-verified that gang placement removes the node-time that a default scheduler accumulates on partially scheduled gangs, and showed that a gating strategy lowers worker idle time and attribute-based backend selection reduces mean per-run cost and end-to-end time. Our approach keeps vendor credentials in the owning user's namespace.

We suggest in our work that the two-queue problem is specific to the scheduler, and not an interface problem. Resource-control middleware such as \gls{qrmi} is necessary to support the scheduler but not sufficient to replace it. The scheduler must represent the second queue, arbitrate contention against it, and defer classical placement until the quantum work is near ready. The precision around doing this scheduling is bounded by the transparency of the second queue. Local or reservable devices would remove that uncertainty, and the graph model we describe generalizes to this case and is capable of expressing finer granularity for devices. As quantum resources move from experimental access to routine allocation in HPC centers, workload managers will need to treat hybrid application patterns as first-class citizens rather than as classical jobs that happen to make an API call by way of plugins. We are excited to extend this design from Kubernetes to the Flux Framework for traditional \gls{hpc}.